\documentclass[aps,pra,numerical,showpacs,preprint]{revtex4-2}
\usepackage{graphicx}
\usepackage[fleqn]{amsmath}
\usepackage{amssymb}
\usepackage{amsfonts}
\usepackage{babel}
\usepackage[utf8]{inputenc}
\usepackage{graphicx, float}
\usepackage{rotating}
\usepackage[normalem]{ulem}
\usepackage{color}

\def\bm#1{\hbox{\boldmath$#1$\unboldmath}}

\usepackage{bbold}

\newcommand*\xbar[1]{%
	\hbox{%
		\vbox{%
			\hrule height 0.5pt % The actual bar
			\kern0.5ex%         % Distance between bar and symbol
			\hbox{%
				\kern-0.1em%      % Shortening on the left side
				\ensuremath{#1}%
				\kern-0.1em%      % Shortening on the right side
			}%
		}%
	}%
}

\begin{document}

\title{Symmetry generators and quantum numbers for
fermionic circularly symmetric motion}

\author{V. B. Mendrot}
\affiliation{São Paulo State University (UNESP), School of Engineering and Sciences, Guaratinguetá, Brazil}

\author{A. S. de Castro}
\affiliation{São Paulo State University (UNESP), School of Engineering and Sciences, Guaratinguetá, Brazil}

\author{P. Alberto}
\affiliation{CFisUC, Physics Department, University of Coimbra,
	P-3004-516 Coimbra, Portugal}

\date{\today}


%Keywords: Dirac equation, symmetries, circular symmetries, spin symmetry

\begin{abstract}
\noindent

The planar dynamics of spin-1/2 quantum relativistic particles is important for several physical systems.
In this paper we derive, by a simple method, the generators for
the continuous symmetries of the 3+1 Dirac equation for planar motion, when there is circular symmetry, i.e.,
the interactions depend only on the radial coordinate. We consider	 a general set of potentials with different Lorentz structures.
These generators allow for several minimal complete sets of commuting observables and their corresponding quantum numbers. We show how they can be used to label the general eigenspinors for this problem.
We also derive the generators of the spin and pseudospin
symmetries for this planar Dirac problem, which arise when
the vector and scalar potentials have the same magnitude and tensor potential and the space components of the four-vector potential are absent.
We investigate the associated energy degeneracies and compare them to the known degeneracies in the spherically symmetric 3+1 Dirac equation.
\end{abstract}

\maketitle

\section{Introduction}
\label{Sec:Introduction}
There are various physical systems in which fermions effectively move in a plane. A prominent example in condensed matter physics is graphene, a two-dimensional crystal composed of a single layer of carbon atoms arranged in a planar lattice of densely packed hexagonal rings (honeycomb structure) \cite{novoselov}. An example of application of the 3+1 dimensional Dirac equation for planar systems is obtained can be seen here by employing a chiral gauge theory that results in a (3+1)-dimensional Dirac hamiltonian for massless fermions in graphene, taking into account the four degrees of freedom of the problem, two in each sub-lattice \cite{PhysRevLett.98.266402,oliveiraVortexGapGeneration2011}, which is more adequate to models in which there are breaking of the sub-lattice symmetry, allowing transitions that mix the two species of fermions obtained for each Dirac point. The investigation of the relativistic planar motion of spin-1/2 fermions in a (3+1)-dimensional setup is therefore of significant interest.

In this study, we describe the symmetries of the 3+1 Dirac equation
when the motion is restricted to the $xy$-plane and when the interactions are circularly symmetric,
i.e., they depend only on the radial coordinate. To accomplish this, we use the general framework of the 3+1 Dirac Hamiltonian in cylindrical coordinates.
This has been previously discussed in other publications and, in particular, we use the formalism of ref.~\cite{de2021spin}.
We believe this formalism is better suited to the main goal of this
work because it maintains a close relationship with the spherically symmetric one for Dirac spinors and thus
facilitates the comparison between the features of circularly symmetric Dirac Hamiltonian and its spinors and the spherically symmetric ones.

The work is organized as follows. In section 2, we present the 3+1 Dirac
equation with a variety of interactions, namely, four-vector, tensor and
scalar, with respect to Lorentz transformations. Restrictions are then made to
planar motion and later to circularly symmetric motion. In section 3, a simple
method for deriving symmetry generators is developed, allowing one to find a set
of observables and study its quantum numbers, from which a subset will later be
chosen to form a complete set of mutually commuting observables, guiding the
building of the respective spinor. In section 4, specific attention is given to spin symmetry,
which arises when the tensor interaction and the space part of the four-vector potential are absent and the vector and scalar
potentials are equal in shape and strength up to a constant. Section 5 shall
be devoted to the pseudospin symmetry, which can be simply obtained from
spin symmetry. In Section 6 we revisit the spinor and the radial
equations. Finally, in section 7 the conclusions are drawn.

\section{Dirac equation} \label{direq}
	
	Let us examine a comprehensive spin-1/2 single-particle problem that encompasses
	four-vector, tensor, and scalar interactions. The time-independent Dirac
	equation in 3+1 dimensions is expressed as	
	\begin{gather}\label{diracEq}
		H\Psi=\varepsilon\Psi,
	\end{gather}
	\noindent where the Hamiltonian is ($\hbar=c=1$)	
	\begin{gather}\label{Hdirac}
		H=\bm{\alpha}\cdot(\bm{p} - \bm{V}) + i\beta\bm{\alpha}\cdot\bm{U} + \beta(m + V_s) + V_v,
	\end{gather}
	\noindent and $\varepsilon$ is the total energy of the state. The interactions are $V^\mu=(V_v,\bm{V})$, $\bm{U}$ and $V_s$, standing for the four-vector, tensor and scalar potentials, respectively. The spinor $\Psi$ has four components. The matrices $\bm{\alpha}$ and $\beta$ obey the algebra $\left\lbrace \alpha_i,\alpha_j\right\rbrace=2\delta_{ij}$, $\left\lbrace \alpha_i,\beta\right\rbrace=0$ and $\beta^2=\mathbb{1}$.
A useful identity for $\bm{\alpha}$ is
	\begin{gather}\label{alphaproperty}	(\bm{\alpha}\cdot\bm{A})(\bm{\alpha}\cdot\bm{B})=\bm{A}\cdot\bm{B}+i(\bm{A}\times\bm{B})\cdot\bm{\Sigma},
	\end{gather}
	\noindent where $\bm{\Sigma}$ is the 4$\times$4 version of the spin matrix.
In this work we choose to work with the standard representation of these matrices \cite{strange}:
	\begin{gather}
		\bm{\alpha}=\left(\begin{matrix}
			0 & \bm{\sigma}\\
			\bm{\sigma} & 0\\
		\end{matrix} \right) \quad\text{ , }\quad \beta= \left(\begin{matrix}
			\mathbb{1}_2 & 0\\
			0 & - \mathbb{1}_2\\
		\end{matrix} \right)\quad\text{,}\quad \bm{\Sigma}= \left(\begin{matrix}
		\bm{\sigma} & 0\\
		0 & \bm{\sigma}\\
	\end{matrix} \right),
	\end{gather}
	\noindent in which $\bm{\sigma}=(\sigma_1,\sigma_2,\sigma_3)$ is the vector of Pauli matrices and $\mathbb{1}_2$ is the $2\times 2$ unit matrix.
	
	Next we define orthogonal projection operators onto the upper and lower two-component spinors
	\begin{gather}\label{proj}
		P_\pm=\frac{\mathbb{1}\pm\beta}{2}.
	\end{gather}	
	\noindent Besides the properties common to all projectors, namely $(P_\pm)^2=P_\pm$, $P_\pm P_\mp=0$ and $P_+ + P_- = \mathbb{1}$, (\ref{proj}) also obeys: $P_\pm\beta=\pm P_\pm$ and $P_\pm\; \bm{\alpha}=\bm{\alpha}\;P_\mp$. Applying those to the spinors
	
	\begin{gather}
		\Psi=\left(  \begin{matrix}
			\varphi\\
			\chi
		\end{matrix}\right),
	\end{gather}
	
	\noindent we get
	
	\begin{gather}
		P_+\Psi=\Psi_+=\left(  \begin{matrix}
			\varphi\\
			0
		\end{matrix}\right) \text{ , } P_-\Psi=\Psi_-=\left(  \begin{matrix}
			0\\
			\chi
		\end{matrix}\right).
	\end{gather}

	\noindent in which $\varphi$ and $\chi$ stands for the upper and lower components of the Dirac spinor.
	
	Writing the sum and difference potentials $V_\Sigma = V_v + V_s$ and $V_\Delta = V_v - V_s$, whose convenience will be clear from the discussions in sections \ref{simspin} and \ref{simpseudopin}, one can cast the Dirac equation as a set of two coupled equations
	\begin{gather}
		\bm{\alpha}\cdot(\bm{p} - \bm{V} - i\bm{U})\Psi_+=(\varepsilon - V_\Delta + m)\Psi_-, \label{psi+}\\
		\bm{\alpha}\cdot(\bm{p} - \bm{V} + i\bm{U})\Psi_-=(\varepsilon - V_\Sigma - m)\Psi_+.\label{psi-}
	\end{gather}
	\noindent Since our aim is to study planar motion, we impose that the potentials are such that there is no dynamics along the $z$-axis. Using  cylindrical coordinates ($\rho$, $\phi$, $z$) and choosing the Coulomb gauge, $\bm{\nabla}\cdot\bm{V}=0$, that constraint means that the potentials for planar motion must be
	\begin{gather}
		\bm{V}=V_\phi (\rho)\bm{\hat{\phi}}\;,\;
		\bm{U}=U_{\rho}(\rho,\phi)\bm{\hat{\rho}}\;,\;
		V_\Sigma=V_\Sigma(\rho,\phi)\;,\;
		V_\Delta=V_\Delta(\rho,\phi).
	\end{gather}
	\noindent In this scenario, a convenient commutation relation is $\left[\bm{\Sigma},\bm{\alpha}\cdot\bm{p}\right]=2\gamma^5 \bm{p}$, where $\gamma^5$ is the chiral operator
	\begin{gather}
		\gamma^5= \left(\begin{matrix}
			0 & \mathbb{1}_2\\
			\mathbb{1}_2 & 0\\
		\end{matrix} \right).
	\end{gather}
	 To obtain decoupled equations for $\Psi_+$ and $\Psi_-$  we use the operator method of ref. \cite{alberto2005tensor} adapted to the cylindrical case.
We will present now the procedure for obtaining the equation for $\Psi_+$.

Let us multiply (\ref{psi+}) by $\bm{\alpha}\cdot(\bm{p} - \bm{V} + i\bm{U})$ by the left and substitute (\ref{psi-}) in the resulting equation:
	\begin{gather}\label{OO}
		\begin{aligned}
			\bm{\alpha}\cdot(\bm{p} - \bm{V} + i\bm{U})\bm{\alpha}\cdot(\bm{p} - \bm{V} - i\bm{U})&\Psi_+= -\bm{\alpha}\cdot\left(\bm{p}V_\Delta\right)\Psi_-
		\\[5pt]
			& + (\varepsilon - V_\Delta + m)(\varepsilon - V_\Sigma - m)\Psi_+.
		\end{aligned}			
	\end{gather}
	\noindent We use then the identity (\ref{alphaproperty}) to obtain
	\begin{gather}\label{OO2}
		\begin{aligned}[t]
			\bm{\alpha}\cdot&(\bm{p} - \bm{V} + i\bm{U})\bm{\alpha}\cdot(\bm{p} - \bm{V} - i\bm{U})\Psi_+=\left\lbrace \bm{p}^2 - 2\frac{\Sigma_z V_\phi + U_\rho}{\rho}\right. \Sigma_zL_z
		\\[10pt]		
			&\left.  + (\Sigma_z V_\phi + U_\rho)^2 - \frac{1}{\rho}\frac{\partial}{\partial \rho}\rho(\Sigma_z V_\phi + U_\rho) - \frac{1}{\rho}\Sigma_z(L_zU_\rho) + 2U_\rho \Sigma_\phi p_z \right\rbrace  \Psi_+,
		\end{aligned}
	\end{gather}
	\noindent where $L_z=-i\partial/\partial\phi$ is the $z$-component of the orbital angular momentum operator, which satisfies the commutation relation $[L_z,\bm{\alpha}\cdot\bm{p}]=\bm{\alpha}\cdot\bm{p}\Sigma_z$, and has the normalized eigenfunctions $\Phi_l=\frac{1}{\sqrt{2\pi}}\,e^{il\phi}$, with eigenvalues $l=0,\pm 1,\pm 2,...$.
	Now we distinguish two cases to further develop (\ref{OO}).

If $\varepsilon - V_\Delta + m = 0$, then
	\begin{gather}\label{2ord0}
	\bm{\alpha}\cdot(\bm{p} - \bm{V} + i\bm{U})\bm{\alpha}\cdot(\bm{p} - \bm{V} - i\bm{U})\Psi_+=0;
	\end{gather}
\noindent if $\varepsilon - V_\Delta + m \neq 0$, we can evaluate the first term in the right hand side of (\ref{OO}) by isolating $\Psi_-$ in (\ref{psi+}) and substituting in it, yielding
	\begin{gather}
			\bm{\alpha}\cdot(\bm{p}V_\Delta)\Psi_-=\begin{aligned}[t]
				&\frac{1}{\varepsilon - V_\Delta + m}\left\lbrace\left[\frac{\partial V_\Delta}{\partial \rho} + \frac{\Sigma_z(L_zV_\Delta)}{\rho}\right] \right.
			\\[10pt]
				&\left. \times \left[\frac{\Sigma_zL_z}{\rho} - \frac{\partial}{\partial \rho} - \Sigma_\phi p_z - (\Sigma_z V_\phi + U_\rho)\right]  \right\rbrace \Psi_+.
			\end{aligned}
	\end{gather}
	\noindent Then the resulting equation for $\Psi_+$ is
	\begin{gather}\label{2ord}
		\mathcal{H}\Psi_+=0,
	\end{gather}
	\noindent where
	\begin{gather}\begin{aligned} \label{2ordOp}
			 &\mathcal{H}=\bm{\alpha}\cdot(\bm{p} - \bm{V} + i\bm{U})\bm{\alpha}\cdot(\bm{p} - \bm{V} - i\bm{U}) - \left( \varepsilon - V_\Sigma - m \right)\left( \varepsilon - V_\Delta + m \right)
	\\[10pt]	
			 &+ \frac{1}{\varepsilon - V_\Delta + m}\left[\frac{\partial V_\Delta}{\partial \rho} + \frac{\Sigma_z(L_zV_\Delta)}{\rho}\right] \left[\frac{\Sigma_zL_z}{\rho} - \frac{\partial}{\partial \rho} - \Sigma_\phi p_z - (\Sigma_z V_\phi + U_\rho)\right].
	\end{aligned}\end{gather}
	\noindent In the absence of the potentials, equation (\ref{2ord}) reduces to the Klein-Gordon equation for all the components of $\Psi_+$. Equation (\ref{2ord0}) is already a Klein-Gordon equation (with vector and scalar couplings) subject to the condition $\varepsilon - V_\Delta + m = 0$. Actually, that equation can be obtained from (\ref{2ord}) by taking the limit $V_\Delta\rightarrow \varepsilon-m$, since, from (\ref{psi+}), one sees that the singularity in the equation (\ref{2ord}) can be removed. Therefore, the first case is included as particular case by the second case, so that there is no need to refer to it separetely.
	
	If one now restricts the potentials even further, so that the motion is also circularly symmetric (analogously to spherically symmetric systems), the following impositions are to be made:
		\begin{gather}\label{condcirc}
		\bm{V}=V_\phi (\rho)\bm{\hat{\phi}}\;,\;
		\bm{U}=U_{\rho}(\rho)\bm{\hat{\rho}}\;,\;
		V_\Sigma=V_\Sigma(\rho)\;,\;
		V_\Delta=V_\Delta(\rho).
	\end{gather}
	\noindent Because of these restrictions, $L_z$ now obeys the commutation relation
	\begin{gather}\label{Lzalpha}
		[L_z,\bm{\alpha}\cdot(\bm{p}-\bm{V}\pm i\bm{U})]=\bm{\alpha}\cdot(\bm{p}-\bm{V}\pm i\bm{U})\Sigma_z.
	\end{gather}
	\noindent By conditions (\ref{condcirc}), the term $(L_zV_\Delta)$ in (\ref{OO2}) vanishes, and the operator (\ref{2ordOp}) reduces to
		\begin{gather}\begin{aligned} \label{OO2ordcirc}
			\mathcal{H} = \bm{\alpha}\cdot(\bm{p} - &\bm{V} + i\bm{U})\bm{\alpha}\cdot(\bm{p} - \bm{V} - i\bm{U}) - \left( \varepsilon - V_\Sigma - m \right)\left( \varepsilon - V_\Delta + m \right)
		\\[10pt]	
			& + \frac{1}{\varepsilon - V_\Delta + m}\frac{d V_\Delta}{d \rho}\left[\frac{\Sigma_zL_z}{\rho} - \frac{\partial\ }{\partial\rho} - \Sigma_\phi p_z -  (\Sigma_z V_\phi + U_\rho)\right].
	\end{aligned}\end{gather}
	\noindent Following ref. \cite{de2021spin}, we take $p_z\Psi=0$. Then, the term which involves $\Sigma_\phi$ in (\ref{OO2ordcirc}) also vanishes.
	\section{Symmetries of the Dirac Hamiltonian} \label{symgen}

    \subsection{General definitions}
	
In this work we are mainly interested in continuous symmetries of the Dirac Hamiltonian when there is circular symmetry, \textit{i.e.}, we  are
looking for operators $\mathcal{O}$ such that
	\begin{gather}
		[H,\mathcal{O}]=0 \quad,\quad \mathcal{O}^{\dagger}=\mathcal{O},
	\end{gather}
	\noindent where $H$ is the Dirac Hamiltonian (\ref{Hdirac}) with the circular symmetric potentials (\ref{condcirc}),
	\begin{gather}
\begin{aligned}
\label{Hdirac_circsym}
		H=&\bm{\alpha}\cdot(\bm{p} - \bm{V}) + i\beta\bm{\alpha}\cdot\bm{U} + \beta[ m+ +\frac{1}{2}(V_\Sigma-V_\Delta)] +\frac12(V_\Sigma+V_\Delta)\\
&=\bm{\alpha}\cdot(\bm{p} - V_\phi (\rho)\bm{\hat{\phi}}) + i\beta\bm{\alpha}\cdot \bm{\hat{\rho}}\,U_{\rho}(\rho) + \beta m +V_\Sigma(\rho)P_++V_\Delta(\rho)P_- \ .
\end{aligned}
	\end{gather}

The operator (\ref{2ordOp}) is

	\begin{gather}\begin{aligned} \label{OO2ordcirc2}
			\mathcal{H} = \bm{\alpha}\cdot(\bm{p} - \bm{V} + i\bm{U})\bm{\alpha}\cdot(\bm{p}& - \bm{V} - i\bm{U}) - \left( \varepsilon - V_\Sigma - m \right)\left( \varepsilon - V_\Delta + m \right)
		\\[10pt]	
			& + \frac{1}{\varepsilon - V_\Delta + m}\frac{d V_\Delta}{d \rho}\left[\frac{\Sigma_zL_z}{\rho} - \frac{\partial\ }{\partial\rho} -  (\Sigma_z V_\phi + U_\rho)\right].
	\end{aligned}\end{gather}
To make this task easier, we build the symmetry generators by studying the symmetries of the simpler equations for the Dirac spinor projections $\Psi_{\pm}$ . By identifying the symmetry-related infinitesimal variations of the projected spinors, and using their coupled equations, one can derive the overall variation of the spinor and thus the corresponding symmetry generator by the relation $\delta\Psi=\delta\Psi_+ + \delta\Psi_-=-i\epsilon\mathcal{O}\Psi$, where $\epsilon$ is an infinitesimal parameter. To be more precise,
we use equation (\ref{2ord}) to find symmetry generators for $\Psi_+$, and then, by means of equations (\ref{psi+}) and (\ref{psi-}), we find the corresponding generator for $\Psi_-$, leading to the overall generator for $\Psi$.

%$\bm{\mathcal{O}}$
%-i\bm{\varepsilon}\cdot\bm{\mathcal{O}}\Psi
If the generators form a vector operator, as, for instance, the generators of SU(2), we shall use the notation $\bm{\mathcal{O}}$ and the infinitesimal parameters will be denoted as a vector $\bm{\epsilon}$.

We say that any general operator $\mathcal{U}$ which satisfies
	\begin{gather} \label{condicoes}
		[\mathcal{H},\mathcal{U}]=0 \quad,\quad \mathcal{U}^{\dagger}=\mathcal{U}
	\end{gather}
	\noindent where $\mathcal{H}$ is given by (\ref{OO2ordcirc2}), is a symmetry generator of $\Psi_+$. In the case of continuous symmetries, the variation of $\Psi_+$ after being operated by the infinitesimal transformation $I-i\epsilon \mathcal{U}$, where $\epsilon$ is an infinitesimal parameter, is given by $\delta\Psi_+=\Psi^{'}_+-\Psi_+=-i\epsilon\mathcal{U}\Psi_+$, where the primed spinor is the transformed one.
%and $\epsilon\mathcal{U}=\sum_{i=1}^{3}\epsilon_i\mathcal{U}_i$ is the linear combination of the three symmetry generators $\mathcal{U}_i$, each one related to an %independent infinitesimal parameter $\epsilon_i$.
	
	Because this is a symmetry transformation, the transformed spinors obey equations (\ref{psi+}) and (\ref{psi-}), which means that, considering the linearity of the equations, one can write
	\begin{gather}
			\bm{\alpha}\cdot(\bm{p} - \bm{V} - i\bm{U})\delta\Psi_+=(\varepsilon - V_\Delta + m)\delta\Psi_-, \label{delpsi+}
	\\
			\bm{\alpha}\cdot(\bm{p} - \bm{V} + i\bm{U})\delta\Psi_-=(\varepsilon - V_\Sigma - m)\delta\Psi_+.\label{delpsi-}
	\end{gather}
	Now there are two possibilities on how to obtain $\delta\Psi_-$ from a known $\delta\Psi_+$: if $\varepsilon - V_\Sigma - m \neq 0$, equation (\ref{delpsi-}) becomes
	\begin{gather}
			\bm{\alpha}\cdot(\bm{p} - \bm{V} + i\bm{U})\delta\Psi_-=(\varepsilon - V_\Sigma - m)\delta\Psi_+=-i(\varepsilon - V_\Sigma - m)\epsilon\mathcal{U}\Psi_+.
	\end{gather}
	\noindent Here we must rewrite the right hand side in order to have $(\varepsilon - V_\Sigma - m)\Psi_+$ to the rightmost position. Then, it can be substituted by (\ref{psi-}), providing, by comparison, an expression to $\delta\Psi_-$; now when $\varepsilon - V_\Delta + m\neq0$, equation (\ref{delpsi+}) simply yields
	\begin{gather}
		\delta\Psi_-=\frac{-i}{\varepsilon - V_\Delta + m}	\bm{\alpha}\cdot(\bm{p} - \bm{V} - i\bm{U})\;\epsilon\mathcal{U}\Psi_+.
	\end{gather}
	\noindent In both cases, the resulting generators must be the same.
	
	Before proceeding to find the symmetry generators, we note that if the symmetry generator candidates for $\Psi_+$ do not have any differential operators in their structure, i.e., $[\mathcal{U},\bm{p}]=0$, the complete generator can be universally determined based on its commutation relation with $\alpha_i$: if it commutes, then $\delta\Psi=-i\epsilon\mathcal{U}\Psi$; if it anticommutes, it follows that $\delta\Psi=-i\epsilon\mathcal{U}\beta\Psi$, because $\beta\Psi_-=-\Psi_-$ and $\beta\Psi_+=\Psi_+$. If none of these are satisfied, the standard procedure described before must be employed.

    \subsection{Generators for circularly symmetric Hamiltonians}

 	We will be looking for spinors with definite parity for treating the planar and circularly symmetric motion. The Dirac equation is covariant under parity transformation \cite{greinerRQM}, yielded by the operator  $P=\lambda_P \beta P_0$, $|\lambda_P|=1$, $P_0\Psi(\bm{r},t)=\Psi(-\bm{r},t)$, and the transformation of the potentials, in the passive view \cite{wachter}, is: $\bm{V} \rightarrow -\bm{V}$, $\bm{U} \rightarrow -\bm{U}$, $V_s \rightarrow V_s$ and $V_v \rightarrow V_v$.
	
	We now proceed to construct the generators. By simple inspection, it is verified that $\Sigma_z$ satisfies the conditions (\ref{condicoes}) for the most general planar motion. Morever, since it anti-commutes with both $\alpha_x$ and $\alpha_y$ --- with the condition $p_z\Psi=0$, $\alpha_z$ will not appear in the Hamiltonian ---, $\beta\Sigma_z$ commutes with $\alpha_x$ and $\alpha_y$. Thus, according to what was stated above, and since $\beta$ commutes with $\Sigma_z$, the quantity $\mathcal{S}_z$, defined as
	\begin{gather}\label{Sz}
		\mathcal{S}_z=\beta\Sigma_z=
\left(\begin{matrix}
		\sigma_z & 0\\
		0 & -\sigma_z\\
	\end{matrix} \right) \ ,
	\end{gather}
is a symmetry generator and therefore one can write $\delta\Psi=-i\epsilon\mathcal{S}_z\Psi$.
	\noindent Its eigenvalues and eigenspinors are then
	\begin{gather}\label{Szautval}
		\mathcal{S}_z\Psi_s=s\Psi_s \quad,\quad s=\pm1 \quad,\quad \Psi_s=\left(\begin{matrix}
			g(\rho,\phi)\chi_s \\[5pt]
			f(\rho,\phi)\chi_{-s}
		\end{matrix}\right),
	\end{gather}
	\noindent where $g$ and $f$ are arbitrary functions insensitive to $\mathcal{S}_z$. The $\chi_s$ are eigenspinors of $\sigma_z$
	\begin{gather}
		\chi_s=\left(\begin{matrix}
			\delta_{s,1} \\[5pt]
			\delta_{s,-1}
		\end{matrix}\right).
	\end{gather}
	Although there is a resemblance, $\mathcal{S}_z$ is not the third component of the spin operator. Indeed, if we extend it to a 3-dimensional vector
	\begin{gather}\label{vecScal}
		\bm{\mathcal{S}}=\beta\bm{\Sigma},
	\end{gather}
\noindent it becomes evident from the algebra of its components, $[\mathcal{S}_i,\mathcal{S}_j]=2i\epsilon_{ijk}\Sigma_k$, that this is not an angular momentum operator.

We now restrict our search of generators to circularly symmetric motion. In equation (\ref{OO2ordcirc}), the explicit appearance of $L_z$ suggests that this operator is also a symmetry generator for $\Psi_+$. Indeed, it obeys conditions (\ref{condicoes}). Using the relationship (\ref{Lzalpha}), it is obtained that

\begin{gather}
		\bm{\alpha}\cdot(\bm{p} - \bm{V} - i\bm{U})L_z=(L_z+\Sigma_z)\bm{\alpha}\cdot(\bm{p} - \bm{V} - i\bm{U})
	\end{gather}

\noindent which allows one to find that $\delta\Psi=-i\epsilon\mathcal{L}_z\Psi$,  where
	\begin{gather}
		\mathcal{L}_z=L_z + P_-\Sigma_z=
\left(\begin{matrix}
		\mathbb{1}_2 L_z & 0\\
		0 & \mathbb{1}_2 L_z+\sigma_z\\
	\end{matrix} \right) \,.
	\end{gather}
	\noindent Since $[\mathcal{S}_z,\mathcal{L}_z]=0$, its eigenspinors can be common to $\mathcal{S}_z$ and furthermore they should have definite parity.
Therefore, these can written as
	\begin{gather}
\label{Lzautval}
		\mathcal{L}_z\Psi_{ls}=l\Psi_{ls} \quad,\quad l=0,\pm1,\pm2,\pm3,... \quad,\quad \Psi_{ls}=\left(\begin{matrix}
			g(\rho)h_{l,s}
		\\[10pt]
			f(\rho)h_{l+s,-s}
		\end{matrix}\right),
	\end{gather}
	\noindent where $g$ and $f$ denote arbitrary functions insensitive to $\mathcal{L}_z$ and $\mathcal{S}_z$, and $h_{l,s}$ are the spinorial circular harmonics, defined by
	\begin{gather}\label{circharm}
		h_{l,s}=\Phi_{l}\,\chi_s \ ,
	\end{gather}
where $\Phi_{l}$ is the eigenfunction of $L_z$ defined in Section \ref{direq}.

	\noindent We take note of the following properties of (\ref{circharm}):
	\begin{gather}
		\int_{0}^{2\pi}d\phi\; h^{\dagger}_{l',s'}h_{l,s}=\delta_{s's}\delta_{l'l},\\[5pt]
		\sigma_\rho h_{l,s}=h_{l,-s}\,,\quad{\rm where}\quad \sigma_\rho=\bm{\sigma }\cdot\hat{\bm{\rho}}
 =\left(
      \begin{array}{cc}
          0 & e^{-i\phi } \\
      e^{+i\phi } & 0%
      \end{array}%
\right) .
	\end{gather}
	\noindent It should be emphasized that although $l$ denotes the eigenvalues of both $\mathcal{L}_z$ and $L_z$, they operate on different objects. The upper component of $\Psi$ --- an object with two components --- is an eigenstate of $L_z$ with eigenvalue $l$, but the lower component has a different $L_z$ eigenvalue,
$l+s$. However, the whole spinor $\Psi$ --- a four-component object --- is an eigenstate of $\mathcal{L}_z$ with eigenvalue $l$.
	
	The new symmetry generator also has not an angular momentum algebra. If we extend it to $\bm{\mathcal{L}}=\bm{L}+P_-\bm{\Sigma}$, the components obey $[\mathcal{L}_i,\mathcal{L}_j]=i\varepsilon_{ijk}L_k$.
	
	From the two generators already determined, it is possible to build additional generators based on then. An interesting observation is that the third component of the total angular momentum can be written in terms of $\mathcal{S}_z$ and $\mathcal{L}_z$
	\begin{gather}\label{Jz}
		J_z=L_z+\frac{\Sigma_z}{2}=\mathcal{L}_z + \frac{\mathcal{S}_z}{2}.
	\end{gather}
	\noindent Thus $J_z$ is also a symmetry generator. Eigenvalues and possible eigenfunctions, again requiring they have definite parity and be simultaneous eigenstates of $\mathcal{S}_z$ and $\mathcal{L}_z$ --- possible since $[\mathcal{S}_z,J_z]=[\mathcal{L}_z,J_z]=0$ ---, are
	\begin{gather}
		J_z\Psi_{m_j,s}=m_j\Psi_{m_j,s} \quad,\quad m_j=l+\frac{s}{2} \quad,\quad \Psi_{m_j,s}=\left(\begin{matrix}
			g(\rho)h_{m_j-s/2,s} \\[5pt]
			f(\rho)h_{m_j+s/2,-s}
		\end{matrix} \right),
	\end{gather}
	\noindent where the functions $g$ and $f$ are the same as in Eq.~(\ref{Lzautval}). The eigenvalue $m_j$ can take the values $\pm 1/2, \pm 3/2, \ldots$ .
	
	Finally, we are also able to build the spin-orbit symmetry generator
	\begin{gather}\label{K}
		K=\mathcal{S}_zJ_z=\beta\left( L_z\Sigma_z + \frac{1}{2}\right) .
	\end{gather}
	\noindent Its eigenvalues and possible eigenfunctions, choosing them to be simultaneous eigenfunctions of $J_z$ --- due to $[J_z,K]=0$ --- and having definite parity, are
	\begin{gather}\label{spinorKJz}
		K\Psi_{k,m_j}=k\Psi_{k,m_j} \quad,\quad k=sm_j=ls+\frac{1}{2} \quad,\quad \Psi_{k,m_j}=\left(\begin{matrix}
			g(\rho)h_{m_j-k/2m_j,\,k/m_j} \\[5pt]
			f(\rho)h_{m_j+k/2m_j,\,-k/m_j}
		\end{matrix}\right),
	\end{gather}
	\noindent where again $g$ and $f$ are the functions defined before. The indexes of spinorial harmonics in this equation were obtained from (\ref{Lzautval}) using the relations $l=m_j-k/2m_j$ and $s=k/m_j$, which in turn results from the choice that the spinor (\ref{spinorKJz}) is to be a simultaneous eigenstate of $K$ and $J_z$ (and of $\mathcal{S}_z$ and $\mathcal{L}_z$). In other words, for a given $k$, $m_j$ can only assume the values $\pm k$ --- this is a similar behavior to the relation between the angular and azimuthal quantum numbers in spherically symmetric systems in non-relativistic theory \cite{griffithsQM}.

	From the four symmetries determined, there are four correspondent eigenvalues which are also quantum numbers: $s$, $l$, $m_j$ and $k$, corresponding respectively to the symmetry generators $\mathcal{S}_z$, $\mathcal{L}_z$, $J_z$ and $K$. Only two of these are independent, because the remaining two can always be written in terms of other two, which is also true, of course, for the generators. The relations are
	\begin{gather}\label{relquantumnumber}
		m_j=l+\frac{s}{2} \quad,\quad k=sm_j.
	\end{gather}
	\noindent Choosing any two of the corresponding symmetry generators, in addition to the $z$-component of the linear momentum operator $p_z$ and the Hamiltonian $H$, one gets a complete set of mutually commuting observables \cite{cohen}, which is enough to completely describe any possible state. Then, the functions $g$ and $f$ may also depend on the selected quantum numbers.

	\section{Spin symmetry}\label{simspin}
	
	Spin symmetry is a spherical relativistic symmetry in which spin-orbit interaction is suppressed for the upper component of the Dirac spinor. Reviews of this symmetry and the related pseudospin symmetry, which is important for describing some nuclei single-particle structure, can be found in references \cite{ginocchio2005relativistic} and \cite{LIANG20151}. Considering back the case of general planar motion, but now without the spatial component of the four-vector interaction and the tensor interaction, \textit{i.e.}, $\bm{V}=\bm{U}=\bm{0}$, we note that the spin-orbit interaction term --- the one involving $L_z\Sigma_z$ --- depends on $V_\Delta$ in such a way that when the potential $V_\Delta$ is constant, this term will vanish, and thus spin symmetry will exist. In this case, (\ref{2ordOp}) reduces to
	\begin{gather}\label{eq2ordsimspin}
		\mathcal{H}=\bm{p}^2 - (\varepsilon - V_\Sigma - m)(\varepsilon - V_\Delta + m).
	\end{gather}
	\noindent Following the procedure derived in section \ref{symgen}, one finds an additional symmetry generator, which is the generalization of (\ref{Sz}). Indeed, the infinitesimal transformation $\delta\Psi_+=-i\bm{\varepsilon}\cdot\bm{\Sigma}\Psi_+$ is a symmetry transformation for (\ref{eq2ordsimspin}), which gives for $\Psi_-$
	\begin{gather}\begin{aligned}[b]
		\delta\Psi_-&=\frac{1}{\varepsilon - V_\Delta + m}\bm{\alpha}\cdot\bm{p}\;\left( -i\bm{\epsilon}\cdot\bm{\Sigma}\Psi_+\right)
	\\[5pt]
		&=-i\bm{\epsilon}\cdot\left(\bm{\alpha}\cdot\bm{p}\frac{\bm{\Sigma}}{\bm{p}^2}\bm{\alpha}\cdot\bm{p}\right) \Psi_-.
	\end{aligned}\end{gather}
	\noindent Finally, the symmetry generator found for the Dirac spinor is
	\begin{gather}\label{O-spin-sym}
		\bm{\mathcal{O}}= \bm{\Sigma}P_+ + \bm{\alpha}\cdot\bm{p}\frac{\bm{\Sigma}}{\bm{p}^2}\bm{\alpha}\cdot\bm{p}P_- .
	\end{gather}
	\noindent Employing (\ref{alphaproperty}), (\ref{vecScal}), and the properties of the projectors, it can also be written as
	\begin{gather}\label{Opsimspin}
			\bm{\mathcal{O}}=\bm{\mathcal{S}} - 2\bm{\hat{p}}(\bm{\mathcal{S}}\cdot\bm{\hat{p}})P_-\;.	
	\end{gather}
	\noindent In this form, it becomes more evident that its third component is exactly the same as (\ref{Sz}), as it should be. Note that, although the generators are formally identical to the spin-symmetric counterpart \cite{alberto2014relativistic}, this behavior of the planar system is different because of the planar motion condition $p_z\Psi=0$.
	
	With the spin symmetry generator available, one can study its algebraic properties, which are given by $[\mathcal{O}_i,\mathcal{O}_j]=2i\epsilon_{ijk}\mathcal{O}_k$ and $\left\lbrace\mathcal{O}_i,\mathcal{O}_j\right\rbrace =2i\delta_{ij}$, with $i$ and $j$ having as possible values $x$,$y$ and $z$, revealing that the spin symmetry generators obey the SU(2) group algebra. The relations of $\mathcal{O}_i$ with the other symmetry generators are found to be
	\begin{gather}
		[\mathcal{O}_i,\mathcal{S}_z]=2i\epsilon_{izk}\mathcal{O}_k, \label{Szcomut}\\[5pt]	
		[\mathcal{O}_i,\mathcal{L}_z]=0, \label{Lzcomut}\\[5pt]
		[\mathcal{O}_i,J_z]=i\epsilon_{izj}\mathcal{O}_j, \label{ComutOiJz}\\[5pt]
		[\mathcal{O}_i,K]=i\varepsilon_{izk}(2\mathcal{O}_kJ_z + \mathcal{O}_z\mathcal{O}_k). \label{ComutOiK}
	\end{gather}
	As referred before, the condition for spin symmetry leads to the  suppression of spin-orbit interaction, namely for the upper component of the Dirac spinor \cite{alberto2014relativistic}. As will be shown, the physical consequence of this is a double-degeneracy in the spectrum of eigenvalues.
	Defining
	\begin{gather}
		\mathcal{O}_{s}\equiv \mathcal{O}_x + is \mathcal{O}_y, \quad s=\pm 1 \ ,
\label{escada}
	\end{gather}
	\noindent we can show that the states $\mathcal{O}_{-s}\Psi_s$, where $\Psi_s$ are given by (\ref{Szautval}), are eigenstates of the symmetry generators.
	
	For the $\mathcal{S}_z$ operator, one gets
		\begin{gather}\begin{aligned}[b]\label{SO}
				\mathcal{S}_z\left(\mathcal{O}_{-s}\Psi_s\right)&=(\mathcal{O}_{-s}\mathcal{S}_z - \left[\mathcal{O}_{-s},\mathcal{S}_z\right])\Psi_s\\[5pt]
				&=-s\left(\mathcal{O}_{-s}\Psi_s\right).	
		\end{aligned}
\end{gather}
	\noindent From (\ref{Szautval}), it can be concluded that $\mathcal{O}_{-s}\Psi_s$ is directly proportional to $\Psi_{-s}$.
Note that this symmetry exists even when there is no circular symmetry.
	
	If circular symmetry exists, then we can consider the effect of the operators $\mathcal{O}_{s}$ on the eigenstates of the corresponding symmetry generators considered in the previous section.

The $\mathcal{L}_z$ operator yields (see eq. (\ref{Lzautval}))
	\begin{gather}\label{LO}
		\mathcal{L}_z\left(\mathcal{O}_{-s}\Psi_{ls}\right)=l\left(\mathcal{O}_{-s}\Psi_{ls}\right).
	\end{gather}
	\noindent Then, $\mathcal{O}_{-s}\Psi_{ls}$ is an eigenstate of $\mathcal{L}_z$, with eigenvalue $l$. Therefore, it is directly proportional to $\Psi_{l,-s}$.

	In the case of the $J_z$ operator, a shifting occurs for the eigenvalue, such that
	\begin{gather}\begin{aligned}[b]\label{JO}
			J_z\left(\mathcal{O}_{-s}\Psi_{m_js}\right)&=(\mathcal{O}_{-s}J_z - \left[\mathcal{O}_{-s},J_z\right])\Psi_{m_js}
		\\[5pt]
			&=(m_j-s)\left(\mathcal{O}_{-s}\Psi_{m_js}\right);
	\end{aligned}\end{gather}

	\noindent The eigenstate $\mathcal{O}_{-s}\Psi_{m_js}$ shows the ladder operator behavior of $\mathcal{O}_{-s}$. Thus, we find that $\mathcal{O}_{-s}\Psi_{m_js}$ is directly proportional to $\Psi_{m_j-s,-s}$.

	Finally, the spin-orbit operator $K$, acting on the state, leads to
	\begin{gather}\begin{aligned}[b]\label{KO}
			K\left(\mathcal{O}_{-s}\Psi_{ks}\right)&=(\mathcal{O}_{-s}K - \left[\mathcal{O}_{-s},K\right])\Psi_{ks}
		\\[5pt]
			&=(-k+1)\left(\mathcal{O}_{-s}\Psi_{ks}\right).
	\end{aligned}\end{gather}
	\noindent The eingenvalue changes in a more drastic manner, by changing the sign of the original eigenvalue, and then shifting it by 1. We conclude that $\mathcal{O}_{-s}\Psi_{ks}$ is directly proportional to $\Psi_{-k+1,-s}$.

	Since $\mathcal{O}_i$ are symmetry generators, it follows that $[\mathcal{O}_s,H]=0$. Consequently, $\Psi_{k,m_j,s}$ and $\mathcal{O}_{-s}\Psi_{k,m_j,s}$ must both be eingestates of $H$, having the same degenerate eigenvalue $\varepsilon$, corresponding to the quantum numbers $k,m_j$ --- and consequently, $s$. It must be reiterated that, due to the second relation in (\ref{relquantumnumber}), a choice of $k$ and $m_j$ uniquely determines $s$, such that its mention usually can be omitted without loss of information. Here, however, we choose to still show it, due to its relation with the operator $\mathcal{O}_{s}$. Furthermore, collecting (\ref{SO}), (\ref{LO}), (\ref{JO}), (\ref{KO}) and the conclusions drawn from it, one finds that
	\begin{gather}
	 \mathcal{O}_{-s}\Psi_{k,m_j,s} \propto\Psi_{-k+1,m_j-s,-s},
	\end{gather}
	\noindent has the same energy eigenvalue as $\Psi_{k,m_j,s}$.

In the case of spin symmetry for potentials with spherical symmetry, there an extra SU(2) symmetry, as described, for instance, in ref.
\cite{alberto2014relativistic}. This comes about because, since there is no spin-orbit interaction, the equation for the upper component the Dirac spinor is invariant under the orbital angular momentum operator $\bm L$. Then, using the same procedure of the previous section an overall generator $\bm{\mathcal{L}}$ is obtained with same structure as $\bm{\mathcal{O}}$ given by (\ref{O-spin-sym}). In our case, because of the restriction to planar motion,
%(the circularly symmetric potentials can be made strong enough to limit the motion to basically the $z=0$ plane)
the only component of $\bm L$ which is relevant is $L_z$ commuting with (\ref{eq2ordsimspin}). This, in turn, becomes the overall general generator $\mathcal{L}_z$ already discussed in the previous section, so there is no new generator related with spin symmetry and circular symmetric potentials.
	
	\section{Pseudospin symmetry} \label{simpseudopin}
	
	As mentioned before, there is another SU(2) symmetry which acts in a similar way as spin symmetry, but now suppressing the spin-orbit coupling in the lower component of the spinor \cite{alberto2014relativistic}. If one considers the second order decoupled equation for $\Psi_-$
	\begin{gather}\label{2ord_psi-}
		\mathcal{H}'\Psi_-=0,
	\end{gather}
where the circularly symmetric Hamiltonian $\mathcal{H}'$ is given by
		\begin{gather}\begin{aligned} \label{OO2ordcirc3}
			\mathcal{H}' = \bm{\alpha}\cdot(\bm{p} - \bm{V} - i\bm{U})\bm{\alpha}\cdot(\bm{p}& - \bm{V} + i\bm{U}) - \left( \varepsilon - V_\Sigma - m \right)\left( \varepsilon - V_\Delta + m \right)
		\\[10pt]	
			& + \frac{1}{\varepsilon - V_\Sigma - m}\frac{d V_\Sigma}{d \rho}\left[\frac{\Sigma_zL_z}{\rho} - \frac{\partial\ }{\partial\rho} -  (\Sigma_z V_\phi - U_\rho)\right] \ ,
	\end{aligned}\end{gather}

	\noindent one sees that if $\bm{V}=\bm{U}=0$ and $V_\Sigma$ is constant, a new symmetry generator would be found by construction starting with the transformation $\delta\Psi_-=-i\bm{\varepsilon}\cdot\bm{\Sigma}\Psi_-$. One finds that the pseudospin symmetry generator $\tilde{\bm{\mathcal{O}}}$ can easily be obtained from the spin symmetry generator (and vice-versa), by the transformation \cite{ginocchio2005relativistic}
%	
	%The free Dirac equation is invariant under a charge conjugation transformation, realized by the operator $C=i\beta\alpha_2 C_0$, with $C_0\Psi=\Psi^*$. For it to be a symmetry of the Dirac equation with couplings, the potentials must transform as $\bm{V} \rightarrow -\bm{V}$, $\bm{U} \rightarrow -\bm{U}$, $V_s \rightarrow V_s$ and $V_v \rightarrow -V_v$. Thus, it transforms $V_\Delta$ into $-V_\Sigma$. Transforming the spin symmetry generator (\ref{Opsimspin}) by a charge conjugation, we obtain the pseudospin symmetry generator \cite{alberto2014relativistic}
%	
	\begin{gather}
		\tilde{\bm{\mathcal{O}}}=\gamma^5\bm{\mathcal{O}}\gamma^5.
	\end{gather}
 The study of the second order equation for $\Psi_-$ also leads to another symmetry generator, obtained by the same means from $\mathcal{L}_z$. It is $\tilde{\mathcal{L}}_z=\gamma^5\mathcal{L}_z\gamma^5$. Thus the results of the previous section can be easily mapped to the pseudospin symmetry case.
 Note, however, that pseudospin symmetric bound states exist only for negative energy (charge conjugated) states if the binding potential goes to zero at large distances \cite{PhysRevC.81.064324} , but can exist for positive energy solutions for confining potentials that go to infinity at large distances
 \cite{PhysRevC.87.031301}.
	
	%Employing the same method used to derive which quantum states have degenerate energy, for the pseudospin symmetry one finds that the state
	%\begin{gather}
	%	\mathcal{\tilde{O}}_{-s}\Psi_{k,m_j,s} \propto\Psi_{-k-1,m_j-s,-s},
	%\end{gather}

	%\noindent has the same energy eigenvalue as $\Psi_{k,m_j,s}$

	\section{Spinors and radial equations} \label{spinorradeq}
	
	In this section, for the sake of completeness, we write down the radial equations for the general problem of the 3+1 Dirac equation with circular symmetry for planar motion.

Choosing the Dirac spinor to have definite parity, we choose as the complete set of mutually commuting observables the third component of the total angular momentum $J_z$ (\ref{Jz}) and the spin-orbit operator $K$ (\ref{K}), besides the Hamiltonian $H$ (\ref{Hdirac}) and the implicitly conserved $z$-component of the linear momentum $p_z$. Thus, the spinor constructed in (\ref{spinorKJz}) (inspired by the one used in \cite{de2021spin}) can be employed. For convenience, we write
	\begin{gather}\label{espinor}
		\Psi_{km_j}=\frac{1}{\sqrt{\rho}}\left(\begin{matrix}
			ig_{k}(\rho)h_{km_j}\\[5pt]
			f_{k}(\rho)h_{-km_j}
		\end{matrix}\right).
	\end{gather}

	The probability density will simply be
	\begin{gather}
		j_0=\Psi^{\dagger}_{km_j}\Psi_{km_j}=\frac{1}{\rho}(\left|g_{km_j}\right|^2 + \left|f_{km_j}\right|^2).
	\end{gather}
	Substituing (\ref{espinor}) in the Dirac equation (\ref{diracEq}), considering the restrictions (\ref{condcirc}) for the motion to be circularly symmetric, the radial equations are obtained
		\begin{gather}
		\dfrac{dg_k}{d\rho} - \dfrac{k}{\rho}g_k + \left(U_{\rho} + \dfrac{k}{m_j}V_\phi\right)  g_k = (m + \varepsilon - V_\Delta)f_k\label{g1}\\[5pt]
		\dfrac{df_k}{d\rho} + \dfrac{k}{\rho}f_k - \left(U_{\rho} + \dfrac{k}{m_j}V_\phi\right) f_k = (m - \varepsilon + V_\Sigma)g_k\label{f1}\ .
	\end{gather}
	Working out the second order equations, the effect of the spin and pseudospin symmetries becomes evident: the spin symmetry condition ($V_\Delta$ is constant) decouples the equation for the upper component radial function $g$, while the pseudospin symmetry condition ($V_\Sigma$ is constant) decouples the equation for the lower component radial function $f$
	\begin{gather}
			\begin{aligned}
				\frac{d^2g_k}{d\rho^2} + \left[-\frac{k(k-1)}{\rho^2} + \frac{dU_\rho}{d\rho} + \frac{2k}{\rho}U_\rho - U^2_\rho + \frac{k}{m_j}\frac{dA_\phi}{d\rho} + \frac{2k^2}{m_j\rho}A_\phi - A^2_\phi\right] g_k
			\\[5pt]
				=-\frac{dV_\Delta}{d\rho}f_k - \left[\varepsilon^2 - m^2 - (\varepsilon+m)V_\Sigma - (\varepsilon-m)V_\Delta + V_\Sigma V_\Delta \right]g_k,
			\end{aligned}
		\\[5pt]
			\begin{aligned}
				\frac{d^2f_k}{d\rho^2} + \left[-\frac{k(k+1)}{\rho^2} - \frac{dU_\rho}{d\rho} + \frac{2k}{\rho}U_\rho - U^2_\rho - \frac{k}{m_j}\frac{dA_\phi}{d\rho} + \frac{2k^2}{m_j\rho}A_\phi - A^2_\phi\right]f_k
			\\[5pt]
				=\frac{dV_\Sigma}{d\rho}g_k - \left[\varepsilon^2 - m^2 - (\varepsilon+m)V_\Sigma - (\varepsilon-m)V_\Delta + V_\Sigma V_\Delta \right]f_k.
			\end{aligned}		
	\end{gather}

	One may note here that, due to a lapse, the coupling terms, proporcional to the derivatives of $V_\Delta$ and $V_\Sigma$, are missing in the second order equations for the radial functions in ref. \cite{de2021spin}.
	\section{Concluding remarks} \label{conclusion}	
	In the present work we have developed a simple method to derive symmetry generators for the Dirac equation. By studying a comprehensive spin-1/2 single-particle problem involving scalar, vector and tensor couplings, we showed, for the case of planar motion with circular symmetry, how to determine symmetry generators by studying the second order equations for the upper and lower components of the Dirac spinor. This method provided a number of symmetry generators, a subset of which could be chosen to form a complete set of mutually commuting observables to construct the physical states. Thus, the present work provides a complete and extensive framework to handle planar and circularly symmetric motion for the 3+1 Dirac equation, determining a general list of conserved quantities, as well as their correspondent quantum numbers and the relations between them, enabling interested readers to delve deeper in the study of such systems.
	
	With the method presented we were also able to analyze the known spin and pseudospin symmetries in a new light, determining its main features for the circularly symmetric scenario, extending our knowledge about them. Furthermore, by casting the generators in a new fashion, we could study in a systematic way how the spin and pseudospin symmetries relate to the other quantum numbers, enabling one to determine exactly which eigenvalues of energy are degenerated as consequence of the symmetries.
	\section{Acknowledgments}	
	The study was financed in part by the Coordenação de Aperfeiçoamento de Pessoal de Nível Superior - Brasil (Capes) - Finance Code 001 and by FCT - Fundação para a Ciência e Tecnologia, I.P. in the
framework of the projects UIDB/04564/2020 and UIDP/04564/2020, with DOI identifiers 10.54499/UIDB/
04564/2020 and 10.54499/UIDP/04564/2020, respectively. PA would like to thank the São Paulo State University, Guaratinguetá Campus, for supporting his stays in its Physics Department. AC would like to thank Universidade de Coimbra for supporting his stays at its Physics Department.
%		
	%\nocite{*} % to test all bib entrys
	\bibliographystyle{unsrt}

%	\bibliography{references_2024-06-04.bib} % file .bib
%	
\end{document}